\documentclass[reprint,aps,superscriptaddress,nofootinbib,groupedaddress,preprintnumbers]{revtex4-1}

\usepackage{amsmath,amsthm,amsfonts}
\usepackage{graphicx}
\usepackage{physics}
\usepackage{hyperref}
\hypersetup{
	colorlinks   = true, 
	urlcolor     = blue, 
	linkcolor    = blue, 
	citecolor    = red   
}

\graphicspath{{figs/}}

\newcommand{\nc}{\newcommand}

\nc{\calR}{{\cal{R}}}
\nc{\calP}{{\cal{P}}}
\nc{\cN}{ {\cal{N}} }

\nc{\Mpt}{M_{_{\rm Pl}}^2}

\usepackage{tikz}
\usetikzlibrary{arrows,shapes}
\usetikzlibrary{trees}
\usetikzlibrary{matrix,arrows} 				
\usetikzlibrary{positioning}				
\usetikzlibrary{calc,through}				
\usetikzlibrary{decorations.pathreplacing}  
\usepackage{pgffor}							

\usetikzlibrary{decorations.pathmorphing}	
\usetikzlibrary{decorations.markings}
\tikzset{
    vector/.style={decorate, decoration={snake}, draw},
	provector/.style={decorate, decoration={snake,amplitude=2.5pt}, draw},
	antivector/.style={decorate, decoration={snake,amplitude=-2.5pt}, draw},
    fermion/.style={draw=black, postaction={decorate},
        decoration={markings,mark=at position .55 with {\arrow[draw=black]{>}}}},
    fermionbar/.style={draw=black, postaction={decorate},
        decoration={markings,mark=at position .55 with {\arrow[draw=black]{<}}}},
    fermionnoarrow/.style={draw=black},
    gluon/.style={decorate, draw=black,
        decoration={coil,amplitude=4pt, segment length=5pt}},
    scalar/.style={dashed,draw=black, postaction={decorate},
        decoration={markings,mark=at position .55 with {\arrow[draw=black]{>}}}},
    scalarbar/.style={dashed,draw=black, postaction={decorate},
        decoration={markings,mark=at position .55 with {\arrow[draw=black]{<}}}},
    scalarnoarrow/.style={dashed,draw=black},
    electron/.style={draw=black, postaction={decorate},
        decoration={markings,mark=at position .55 with {\arrow[draw=black]{>}}}},
	bigvector/.style={decorate, decoration={snake,amplitude=4pt}, draw},
}

\tikzstyle{block} = [draw, rectangle, 
    minimum height=3em, minimum width=6em]

\begin{document}


\title{Tilt and Tensor-to-Scalar Ratio in Multi-Scalar Field Inflation: Non-Sum-Separable Case}

\author{Fereshteh Felegary,$^{1}$ Seyed Ali Hosseini Mansoori,$^{2}$ Tahere Fallahi Serish$^{2}$ and Phongpichit Channuie$^{1,3}$}
\email{fereshteh.felegary@gmail.com,\ fallahi.ta@shahroodut.ac.ir,\ shosseini@shahroodut.ac.ir,\ phongpichit.ch@mail.wu.ac.th}
 
 \affiliation{$^{1}$School of Science, Walailak University,
Nakhon Si Thammarat, 80160, Thailand}        
\affiliation{$^{2}$Faculty of Physics, Shahrood University of Technology, P.O. Box 3619995161 Shahrood, Iran}
\affiliation{$^{3}$College of Graduate Studies, Walailak University,
Nakhon Si Thammarat, 80160, Thailand}

\begin{abstract}
The canonical multi-scalar field inflation where the kinetic and potential terms are sum-separable is ruled out by  the current observations for the chaotic-type potential $V=\sum_{i} \mu_{i} \phi_{i}^{p}$. 
This paper explores the non-sum-separable case to validate the chaotic-type potential in the multi-scalar field, incorporating a linear coupling term between the kinetic and potential terms in the canonical Lagrangian. This coupling influences the slow-roll parameters and also alters our predictions for the spectral index $n_{s}$ and the tensor-to-scalar ratio $r$, which directly depend on those parameters. In fact, compared to standard canonical multi-field inflation, the values of $n_{s}$ and $r$ decrease to levels consistent with the recent Planck+BICEP/Keck constraint.  

\end{abstract}

\maketitle 

\section{Introduction}\label{sec0}

In contemporary cosmology, inflation is the widely accepted framework for understanding the physics of the early universe and the mechanisms behind the formation of large-scale structures. A fundamental prediction of inflation models is that primordial perturbations are nearly scale-invariant, adiabatic, and Gaussian, which are consistent with cosmological observations \cite{akrami2020planck}.

Nonetheless, the precise nature of the field(s) responsible for driving inflation remains unknown.
In the original realization of inflation model, a scalar field or the inflaton is considered to drive the accelerated expansion to achieve the number of e-foldings required for solving the flatness and the horizon issues \cite{Albrecht:1982wi,Linde:1981mu,Starobinsky:1980te,Guth:1980zm}. Moreover, the validity of such inflationary single scalar field models has been tested by observations such as the cosmic microwave background (CMB), large-scale structure, and so on. 

In the context of inflationary cosmology, two key observables are essential for understanding the early universe: the scalar spectral index $n_{s}$ and the tensor-to-scalar ratio $r$. The spectral index quantifies how the amplitude of primordial scalar perturbations varies with scale, indicating the degree to which these perturbations diverge from a perfectly scale-invariant spectrum. Recent observations from Planck have reported the best-fit value of the spectral index as approximately $n_{s} \approx 0.9649 \pm 0.0042$. Meanwhile, the tensor-to-scalar ratio measures the relative strength of primordial gravitational waves (tensor perturbations) compared to density fluctuations (scalar perturbations) in the early universe. The recent observations from Planck and BICEP/Keck during the 2018 observation period \cite{ade2021improved} indicate that the tensor-to-scalar ratio parameter is constrained to $r < 0.036$ with 95$\%$ confidence. 
For example, the chaotic inflation with a potential $V(\phi) \propto \phi^{p}$, even for $p=2/3$, has been ruled out by these observations.

Although the simplest inflation models are based on a single scalar field, models of high energy physics often motivate inflation driven by multiple scalar fields \cite{Wands:2007bd,baumann2011physics,weinberg2008cosmology}.
Many studies have investigated multi-field inflation models, such as double inflation \cite{silk1987,langlois1999,polarski1992,Feng2003,Yamaguchi2001,Polarski1994,Peter1994,Polarski1995,Hodges1990}, N-flation \cite{Dimopoulos2008,Easther2006,Guo2017,Price2015,Ashoorioon2009,Battefeld2007,Kim2006,Piao2006,Kim2006b}, assisted inflation \cite{Liddle1998,Malik1999,Copeland1999,Kaloper2000,Coley2000}, multi-natural inflation \cite{Kim2005,Ben-Dayan2014,Czerny2014,Choi2014,Higaki2014}, and multi-field monodromy inflation \cite{Wenren2014} (See also the multi-field extension of mimetic inflation \cite{Mansoori:2021fjd}.). In these models, each field can play a role in the inflationary dynamics, implying that all fields may be treated as inflationary fields. Additionally, in certain multi-field inflation models such as the modulated reheating scenario \cite{Dvali:2004,Kofman:2003} and the curvaton model \cite{Moroi:2001,Enqvist:2002,Lyth:2002}, there exists a scalar field known as the spectator field. This field does not influence the inflationary dynamics but does impact primordial density fluctuations. Moreover, such multi-field models can yield specific predictions for primordial non-Gaussianity, as discussed in various research articles, for instance  \cite{Ichikawa:2008,Ichikawa:2008b,Enqvist:2013,Fujita:2014,Vennin:2015,Vernizzi:2006,Huang:2009,Sasaki:2006,Seery:2005,Rigopoulos:2006,Battefeld:2007,Yokoyama:2008,Yokoyama:2007uu,Suyama:2010uj,Frazer:2011tg,Kobayashi:2012ba}. In addition, there have been growing interests in the multi-field inflation to enhance the primordial power spectrum at small scales for PBHs formation \cite{Sakharov:1993qh,Randall:1995dj,Garcia-Bellido:1996mdl,Kawasaki:1997ju}.

The predictions for observables, including the spectral index $n_{s}$ and the tensor-to-scalar ratio $r$, have been extensively studied in these multi-field inflation models \cite{Wenren:2014,Langlois:2004,Moroi:2005,Moroi:2005b,Ichikawa:2008,Ichikawa:2008b,Enqvist:2013,Fujita:2014,Vennin:2015,Easther:2005zr}. Similar to chaotic single-field inflation, chaotic-type potentials represented by $V=\sum_{i}\mu_{i}\phi_{i}^{p}$ are still excluded by current Planck constraints \cite{ade2021improved}, even in multi-field configurations \cite{Easther:2005zr,Wenren:2014}, due to the high tensor-to-scalar ratio. Nevertheless, it has been shown in Ref. \cite{Morishita:2022bkr} that the model with the potential $V=\sum_{i}^{3} \mu_{i} \phi_{i}^{2}$ can align with recent observational constraints on $n_{s}$ and $r$ when three fields with a specific hierarchical mass spectrum are present, where two fields act as inflatons and the third serves as a spectator.

On the other hand, in Ref. \cite{HosseiniMansoori:2024pdq}, a novel subset of the K-essence model was recently introduced, featuring a coupling term between the potential function and the kinetic term. This coupling term can affect the slow-roll parameters, potentially altering the values of the spectral index $n_{s}$ and the tensor-to-scalar ratio $r$ for the $V \propto \phi^{p}$ models, even for $p=2/3$, falling entirely within the region determined by current observational results. A similar result was reported for the $\mathbb{T}^2$-inflation model \cite{Mansoori2024,Mansoori2023}, which is equivalent to the model introduced in Ref. \cite{HosseiniMansoori:2024pdq} in the slow-roll approximation.

Building on the findings of \cite{HosseiniMansoori:2024pdq}, in this paper we consider a special multi-field model, which incorporates a linear coupling term between the multi-scalar field potential and the canonical Lagrangian, given by Eq. \eqref{action}. 

Unlike the findings in Ref. \cite{Morishita:2022bkr}, our multi-field framework involves all scalar fields contributing to inflation through several stages. At each stage, only one scalar field undergoes slow roll and subsequently decays, while the others remain static. The next stage of inflation is initiated by the second field before it decays, and this pattern continues. For this scenario to occur, a hierarchical arrangement of the masses of the scalar fields $\Phi^{a}$ is necessary, with the most massive field rolling first, followed by the second most massive, and so forth. Therefore, similar to the single-field scenario \cite{HosseiniMansoori:2024pdq}, we expect that this term changes the slow-roll parameters, thereby potentially altering the values of $n_s$ and $r$ results reported in the sum-separable multi-field canonical models \cite{Easther:2005zr,Wenren:2014}.

The remainder of the paper is organized as follows. In Section \ref{Sec1}, we start by introducing our model and deriving the background equations in FLRW spacetime. Next, we analytically calculate the field dynamics in the slow-roll limit as a function of the number of e-folds and compare these analytical results with numerics. At the beginning of Section \ref{Sec2}, after formulating the quadratic action of scalar perturbations in the spatially-flat gauge, we discuss the model's stability by imposing constraints on its free parameter. In the rest of this section, after calculating the power spectrum of the scalar curvature perturbation using the $\delta N$ formalism \cite{Sasaki:1995aw,Sasaki1995}, we attempt to analytically calculate $n_{s}$ and $r$ as functions of the number of e-folds. Before moving on to the next section, we verify the accuracy of these analytical results against numerical results for a quadratic potential. Section \ref{Sec3} summarizes the results of this work and draws conclusions.

\section{The Model and the background equations}\label{Sec1}
As an extension of the non-sum-separable Lagrangian introduced in Ref. \cite{HosseiniMansoori:2024pdq} to the multi-field case, we consider the following function.
\begin{equation}\label{action}
P(X,V)=f(\Phi^{a})\Big(X - V(\Phi^{a})\Big),
\end{equation}
where $X \equiv -\delta_{ab}\partial_{\mu}\Phi^{a} \partial^{\mu}\Phi^{b}/2$ is the kinetic term and $V$ is a potential function of scalar fields $\Phi^{a}$ ($a=1,2,..,\text{N}$). It is noteworthy that we are working in a curved field space, whose metric is conformally related to that of the flat field space by  $G_{ab}=f(\Phi^{c}) \delta_{ab}$  \cite{Langlois2008,Gong2011,Gong2016,Gong20161}. In our setup, $f$ can be considered a functional of the potential function, defined by
\begin{equation}\label{f}
f(\Phi^{a}) = 1-\frac{\mathcal{K}}{M_{\rm pl}^{4}} V(\Phi^{a}),
\end{equation}
in which $M_{\rm pl}$ is the reduced Planck mass and $\mathcal{K}$ is a dimensionless constant. Notice that as one sets $\mathcal{K}=0$, the $f$ function reduces to the canonical multi-field inflation Lagrangian, where the kinetic and potential terms are sum-separable as $P(X,V)=X-V$ \cite{Easther:2005zr,Wenren:2014}. Obviously, in comparison with the canonical multi-field inflation Lagrangian, there is an additional coupling term between the kinetic and potential terms, given by $X V(\Phi^{a})$. For the single field inflation scenario, one of the authors of this paper has shown in Ref. \cite{HosseiniMansoori:2024pdq} that such a term alleviates the constraints on the inflationary parameters, such as the spectral index $n_{s}$ and the scalar-to-tensor ratio $r$ compared to the standard chaotic inflation. Therefore, we expect that this coupling will also alter the predictions on inflationary parameters in the multi-scalar field inflation. In the remainder of this paper we will investigate the impact of such a term on  the spectral index  and the tensor-to-scalar ratio parameters. For convenience, we set $M_{\text{pl}} = 1$ throughout this paper.

 For Lagrangian \eqref{action}, the background equations of motion in a spatially flat FLRW spacetime are derived by \cite{Langlois2008,Gong2016}

\begin{eqnarray}\label{H2}
 H^{2} &=&  \frac{1}{3}f(\phi^{a})\Big(\frac{\dot{\phi}_{0}^{2}}{2} + V(\phi^{a})\Big),\\
\label{dotH}
\dot{H} &=&  -f(\phi^{a})\frac{\dot{\phi}_{0}^{2}}{2}, \\
\mathcal{D}_{t} \dot{\phi}^{a}&+&3 H \dot{\phi}^a+\frac{ \delta^{ab}}{f(\phi^{c})} \tilde{V}_{,b}=0,\label{eqpdd}
\end{eqnarray}
where $\tilde{V}(\phi^{a})=f(\phi^{a})V(\phi^{a})$ and $\mathcal{D}_{t}$ is a covariant time derivative with $\Gamma_{bc}^a$ being the Christoffel symbol constructed by the field space metric $ G_{ab}=f \delta_{ab}$\footnote{Note that we have used $\mathcal{D}_{t}f=0$ which follows from the definition of the covariant differentiation.}, which is defined as
\begin{equation}\label{dtexp}
\mathcal{D}_{t} A^a\equiv \dot{A}^{a}+ \Gamma_{bc}^{a}\dot{\phi}^{b} {A}^{c},
\end{equation}
in which $\Gamma_{bc}^{a}=(\delta_{b}^{a} f_{,c}+\delta_{c}^{a} f_{,b}-f^{,a} \delta_{bc})/2f$. In addition, $H\equiv\dot{a}/a$ is Hubble parameter in which $a$ is the scale factor, $\phi^{a}$ is the homogeneous part of the scalar field $\Phi^{a}$, and $\dot{\phi}_{0}^{2} \equiv \delta_{ab} \dot{\phi}^{a}\dot{\phi}^{b}$. Moreover, the dot and the comma stand for the derivative with respect to the cosmic time and scalar fields, respectively. In the case of 
 $\mathcal{K}=0$ and $f=1$, the above equations of motion reduce to the standard canonical multifield background dynamics \cite{Easther:2005zr,Wenren:2014}. Utilizing Eq. \eqref{dtexp}, Eq. \eqref{eqpdd} can be rewritten as
 
 \begin{equation}
 \ddot{\phi}^{a} + \Big(3H + \frac{\dot{\phi}^{b} f_{,b}}{f} \Big) \dot{\phi}^{a} - \frac{\delta^{ab} V_{,b}}{f}\Big(1- \frac{\mathcal{K}\dot{\phi}_{0}^{2}}{2}-2 f \Big)=0. \label{phidotH}
 \end{equation}
In the slow-roll approximation, when $\dot{\phi}_{0}^{2} \ll V$ and $\ddot{\phi}^{a}\ll H \dot{\phi}^{a}$, above relations reduce to 
\begin{eqnarray}\label{H2multi}
H^{2}\simeq \frac{1}{3}f(\phi^{a}) V(\phi^{a}),\\
\label{ddotphimulti}
3H \dot{\phi}^{a} \simeq  -\frac{\delta^{ab}V_{,b}(2f-1)}{f}.
\end{eqnarray}
From Eq. \eqref{H2multi}, one deduces that there is an upper bound on the potential, namely $f>0$. This finding also appears to be well  supported by satisfying the Null Energy Condition (NEC) in Eq. \eqref{dotH}, which means that $\dot{H}<0$. Therefore, we can find an upper bound on the potential function, given by $V < 1/\mathcal{K}$, by assuming $ \mathcal{K} > 0$, without loss of generality. 
 
When dealing with inflationary dynamics, it is numerically advantageous to use the number of e-folds $N$, defined by $dN \equiv d\ln(a)=H dt$, as the independent variable for evolving the equation \cite{Li:2012vta}. 
Therefore, by combining the above equations, it becomes straightforward to verify that
\begin{equation}\label{dphidN}
\frac{d \phi^{a}}{dN}=-\frac{\delta^{ab}V_{,b}}{V}\Big(\frac{2f-1}{f^2}\Big).
\end{equation}
Now, making use of  Eqs. \eqref{dotH}, \eqref{H2multi} and \eqref{dphidN}, one obtains the Hubble slow-roll parameter as follows. 
\begin{eqnarray}\label{epsilonH}
\nonumber \epsilon_{H}&\equiv&-\frac{\dot{H}}{H^2}\simeq \frac{1}{2} f \delta_{ab} \frac{d \phi^{a}}{dN} \frac{d \phi^{b}}{dN}\\
&\simeq & \frac{1}{2} \Big(\frac{V_{,a}V^{,a}}{V^{2}}\Big) \frac{(2f-1)^{2}}{f^{3}}.
\end{eqnarray}

As mentioned, in our inflation scenario, we consider multiple scalar fields $\phi^{a}$ $(a = 1, 2, \ldots, n)$ driving inflation in $n$ stages. In fact, similar to N-flation \cite{Dimopoulos2008,Easther2006} in each stage, only one scalar field undergoes slow-roll and then decays, while the others remain frozen. For instance, in $n=2$ case, the first inflationary phase is driven by the $\phi_{1}$, while the other field remains frozen. After $\phi_{1}$ reaches its minimum and its energy dissipates following a few rapid oscillations, the $\phi_{2}$ field sequentially drives the subsequent inflationary phase. 
 
More precisely, in cases with a large mass ratio, specifically when $ \mu_{1} \gg \mu_{2}$ (where $\mu_{i}$ represents the mass of the scalar field $\phi_{i}$),
the evolution of the universe can be distinctly divided into three stages. The first stage is characterized by inflation driven primarily by the $\phi_{1}$-field. The second stage is a non-inflationary phase during which the energy density of the oscillating $\phi_{1}$-field remains greater than that of the $\phi_{2}$-field. The third stage sees inflation again induced by the $\phi_{2}$-field. Consequently, the total number of e-folds throughout these stages is predominantly contributed by the first and third inflationary phases, while the perturbation in the e-folding number during the second stage is negligible. Thus, as a rough approximation, the dynamics can be simplified as the sum of two single-field inflationary phases.



Therefore, the number of e-foldings is achieved by
\begin{equation}\label{Nnew}
N =\sum_{a}N_{a} \simeq \sum_{a}\int_{\phi_{\rm initial}^{a}}^{\phi_{\rm final}^{a}} \Big( \frac{f^{2}}{2f-1}\Big)\Big(\frac{V}{V_{,a}} \Big)d\phi_{a}.
\end{equation}

 
Additionally, for this multi-stage inflation to be realized, we need a hierarchical arrangement of the $\phi^{a}$ masses, such that the most massive field begins rolling first, followed by the second most massive field, and so on.

Regarding Eq. \eqref{Nnew}, in the two-field inflation case with a sum-separable potential $V=\sum_{a=1}^{2} \mu_{a}\phi_{a}^{p}$, the number of e-folds for the first inflationary phase, driven by the scalar field $\phi_{1}$ while $\phi_{2}$ remains nearly frozen (i.e., $\phi_{2}\simeq \phi_{2}^{0}$ where $\phi_{2}^{0}$ is the initial value for $\phi_{2}$.), is given by
\begin{eqnarray}\label{N1formula}
\nonumber N_{1}&=&\frac{\phi_{1}^{2-p}}{8 \mathcal{K} \mu_{1}p(p-2)}\Big(1-{}_{2}F_{1}(1,\frac{2}{p}-1,\frac{2}{p},2 \mathcal{K} \mu_{1} \phi_{1}^{p})\Big)\\
\nonumber &+&\frac{3\phi_{1}^2}{8p}\Big(1-\frac{4 \mathcal{K} \mu_{1}\phi_{1}^p}{3(2+p)}\Big)
- \frac{1}{4 (p-2)p} \Big(\frac{\mu_{2}}{\mu_{1}} (\phi_{2}^{0})^{p}\Big) \\
\nonumber &\times&( \phi_{1}^2 \Big[\phi_{1}^{-p}\Big(3+{}_{2}F_{1}(1,\frac{2}{p}-1,\frac{2}{p},2 \mathcal{K} \mu_{1} \phi_{1}^{p})\Big)\\
\nonumber &+&\mathcal{K} \mu_{1} (p-2)\Big({}_{2}F_{1}(2,\frac{2}{p},\frac{2+p}{p},2 \mathcal{K} \mu_{1} \phi_{1}^{p}))-2\Big)\Big]\\
&+& \mathcal{O}\Big(\Big(\frac{\mu_{2} }{\mu_{1}}\Big)^2 (\phi_{2}^{0})^{2p}\Big).
\end{eqnarray}
where ${}_{2}F_{1}$ is the Gaussian hypergeometric function and we have used the hierarchy of masses, namely $\mu_{2}\ll \mu_{1}$ in the above expansions. 

\begin{figure}
\centering
	\includegraphics[width=0.5\textwidth]{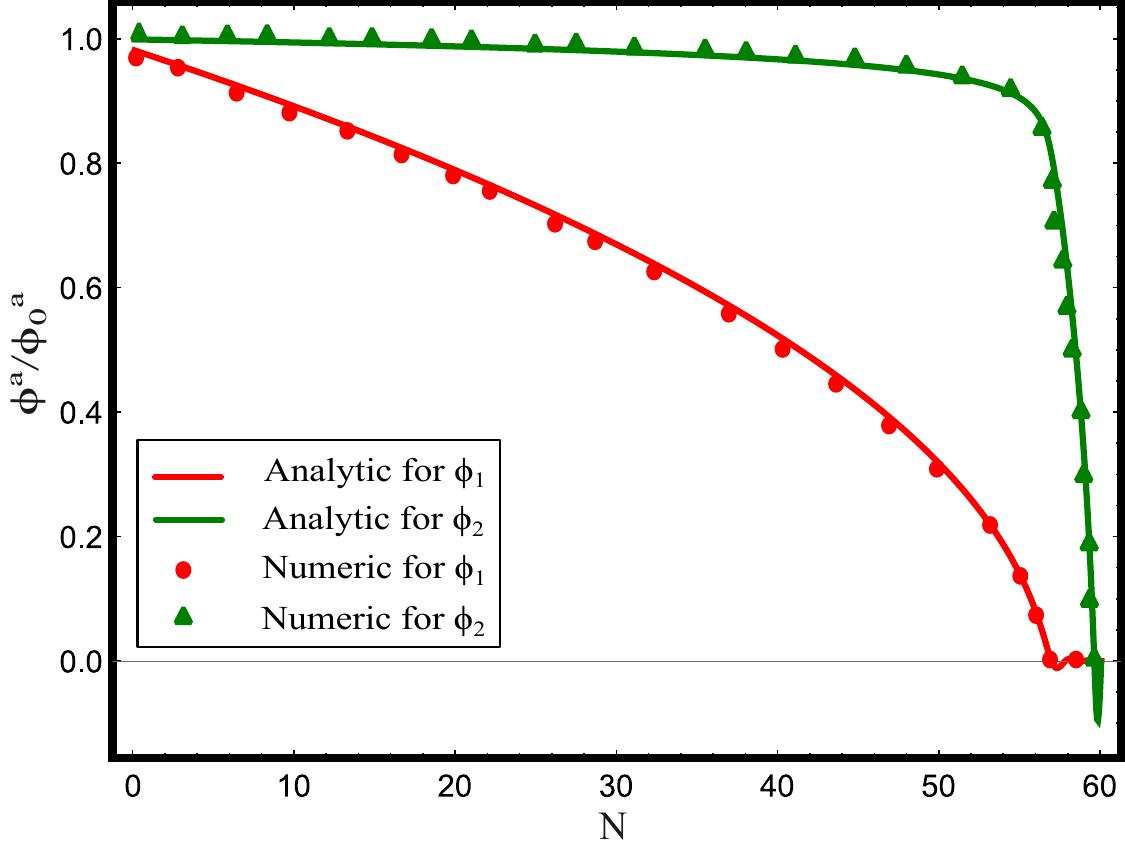}
	\caption{ Evolution of the scalar fields. Clearly, the first inflationary phase is driven by $\phi_{1}$ until $N_{1}=57$, while $\phi_{2}$ remains constant. Once $\phi_{1}$ reaches its potential minimum after $N_{1}$ e-folds, it oscillates rapidly at the bottom of its potential until its amplitude effectively dies out. Then, the next field, $\phi_{2}$, begins its rolling phase until it arrives at its minimum for $N_{2}=3$. The solid red line and filled red circles represent the analytical and numerical evolution of $\phi_{1}$, respectively, while the solid green line and filled green triangles depict the analytical and numerical evolution of $\phi_{2}$.} 
	\label{fig1}
\end{figure}

As mentioned, during the first stage, only the field $\phi_{1}$ rolls down towards its potential minimum for some e-folds $N_{1}$, oscillating rapidly at the bottom of its potential until its amplitude effectively dies out (i.e., $\phi_{1} \sim 0$), and the next field $\phi_{2}$ starts rolling. As a result, the second inflationary phase occurs for
\begin{equation}\label{N2formula}
N_{2}=\frac{\phi_{2}^{2}}{2p}+\mathcal{O}\Big(\Big(\frac{\mu_{2} }{\mu_{1}}\Big)^{2}\Big),
\end{equation}
Due to the complex nature of the hypergeometric function, it is not possible to determine the functions $\phi_{1}$ in reverse as a function of $N_{1}$. However, this can be achieved by selecting smaller values of $\mathcal{K} \mu_{1}$, ensuring that $\beta=\mathcal{K} \mu_{1} \ll \mu_{2}/ \mu_{1}$. In this regime, we can obtain
\begin{eqnarray}\label{anaphi1}
\nonumber \phi_{1}(N) &\simeq & \Big( 2 (N_{1}-N) p\Big)^{\frac{1}{2}}\Big[1 - \frac{\Big( 2 (N_{1}-N) p\Big)^{p}}{2(1+p)}  \beta^{2}\\
& -&
\nonumber \frac{\Big( 2 (N_{1}-N) p\Big)^{\frac{3p}{2}}}{2(2+3p)} \beta^{3}
-\frac{(13+22p+8p^{2})}{2^{3}(1+p)^{2}(1+2p)}\\
&\times &
\Big( 2 (N_{1}-N) p\Big)^{2p}\beta^{4}
+\mathcal{O}(\beta^{5})\Big],
\end{eqnarray}
within the range $N\in \{0, N_{1}\}$. In line with the numerics shown in Fig. \ref{fig1}, we takes the scalar fields value at the CMB pivot scale $k_* = 0.05 \text{Mpc}^{-1}$ given by 
\begin{eqnarray}
&& \phi_{1}(N=0)=\phi_{1}^{0}=\phi_{1}^{\rm CMB}\\
&& \phi_{2}(N=0)=\phi_{2}^{0}=\phi_{2}^{\rm CMB}.
\end{eqnarray}
It should be noted that the accuracy of Eq. \eqref{anaphi1} is maintained as
\begin{equation}\label{limitaccuracy}
\mu_{2} (\phi_{2}^{0})^{p}/\mu_{1} \sim \mu_{2} N_{2}^{p/2}/\mu_{1} \lesssim \mathcal{O}(\sqrt{\mu_{2}/\mu_{1}}).
\end{equation} 
On the other hand, using Eq. \eqref{N2formula} one gets
\begin{equation}\label{anaphi2}
\phi_{2}(N)=\Big(2 p (N_{\rm end}-N)\Big)^{\frac{1}{2}},
\end{equation}
within the range $N \in \{N_{1},N_{\rm end}\}$ where $N_{\rm end}$ is the number of e-foldings at the end of inflation.  Notice that $N_{2}=N_{\rm end}-N_{1}$ and $\phi_{2}(N=N_{1})=\phi_{2}(N=0)=\phi_{2}^{\rm CMB}$, it means that $\phi_{2}$ remains constant in the range $\{0,N_{1}\}$ whereas $\phi_{1}$ is rolling to its potential minimum. 
 
  Although we consider the expansion up to the fifth order here, in some cases, it is necessary to consider higher orders, such as the sixth order, to achieve a strong convergence between analytical and numerical results. 
  
  Interestingly, by taking the limit $\beta \to 0$ in Eqs. \eqref{N1formula} and \eqref{N2formula}, it is easy to verify that $N = N_{1} + N_{2} = (\phi_{1}^2 + \phi_{2}^2) / 2p$. This result aligns with the findings in Refs. \cite{Easther:2005zr,Wenren:2014}.
  
As a concrete example, let us consider a simple choice for the inflation potential by a power of $p=2$. In this case, in order to create two-phase inflation, we need to take $\mu_{1} \sim 10^{-10}$ and $\mu_{2} \sim 10^{-12}$, which manifests the hierarchy of masses, i.e., $\mu_{2}/\mu_{1} \sim 10^{-2}$. As shown in Fig. \ref{fig1}, the background experiences two inflationary phases. The first inflationary phase is driven by the scalar field $\phi_{1}$ while the other field $\phi_{2}$ remains frozen. After $\phi_{1}$ has reached to its minimum and its energy has dissipated after a few rapid oscillations $\phi_{2}$ drives the next inflationary phase each in turn. The numerical data are depicted by colored points, with red filled circles and green filled triangles representing the numerical calculations of Eqs. \eqref{H2}, \eqref{dotH}, and \eqref{eqpdd}. Interestingly, there is perfect agreement between the numerical and analytical results.
 
In our numerical calculations, the variable $N$ denotes the number of e-foldings remaining until the end of inflation. Thus, $N = 0$ marks the start of inflation at the CMB scale, while $N = N_{\rm end}$ signifies its end. 
 Here we consider three values of $N_{\rm end} =60$.
 Notice that the initial values for scalar fields at the CMB scale at $N=0$ are determined by setting the $N_{1}$ and $N_{2}$ values in the analytical relations \eqref{anaphi1} and \eqref{anaphi2}. Here we select $N_{1}=57$ and $N_{2}=3$ so that $N_{\rm end}=N_{1}+N_{2}=60$. Notice that the choice of $N_{2}=3$ satisfies the limit in Eq. \eqref{limitaccuracy}. Specifically, when $p=2$, the numerical and analytical predictions align as long as
 \begin{equation}
 N_{2} \lesssim \sqrt{\frac{\mu_{1}}{\mu_{2}}} \sim 10
 \end{equation}

  \section{Inflationary parameters}\label{Sec2}
  
  This section aims to explore how the coupling $XV$ (or $\beta$ corrections) affects inflationary parameters, such as the spectral index $n_s$ and the tensor-to-scalar ratio $r$. Before we proceed further, let us first discuss the stability of the model through the evolution of scalar perturbations. 
  
  The scalar field $\Phi_a$ can be decomposed into its background part $\phi_a$ and its perturbation $Q_a$ as $\Phi_a = \phi_a + Q_a$. Furthermore, the scalar and tensor perturbations of the metric around the FLRW background metric 
in the spatially-flat gauge, are given by
\begin{equation}
 ds^2=-(1+2 A) dt^2-2 \textbf{a}^2  B_{i} dt dx^{i}+\textbf{a}^2 \delta_{ij} dx^{i} dx^{j}.
\end{equation} 
By substituting the metric and the scalar fields perturbations into the action associated with the Lagrangian \eqref{action}, expanding it to the second order, and integrating out the non-dynamical modes $(A,B)$, the quadratic action in terms of the dynamical mode $Q_{a}$ is obtained as 
\begin{eqnarray}\label{action2}
\nonumber S^{2}&=&\frac{1}{2} \int dt d^{3} x \textbf{a}^3 \Big[f \delta_{ab} \mathcal{D}_{t}Q^{a} \mathcal{D}_{t}Q^{b} -\frac{f}{\textbf{a}^2} \delta_{ab} \partial_{i}Q^{a} \partial_{j}Q^{b}\\
 &-&M_{ab}^2 Q^{a}Q^{b}
 \Big],
\end{eqnarray}
where the effective mass matrix is given by
\begin{equation}
\nonumber M_{ab}^2= \tilde{V}_{;ab}- \mathbb{R}_{acdb} \dot{\phi}^{c}\dot{\phi}^{d}+\Big(\epsilon_{H}-3\Big(\frac{1}{2f-1}\Big)\Big))\dot{\phi}_{a} \dot{\phi}_{b},
\end{equation}
where $\mathbb{R}^{a}_{cdb}$ is the Riemann tensor associated with the above Christoffel symbol \cite{Langlois2008,Gong2016}. 
 
From Eq. \eqref{action2}, we conclude that as $f > 0$, our model is free from ghost and gradient instabilities. This result confirms our earlier discussion following Eq. \eqref{ddotphimulti}. Furthermore, similar to the standard canonical model, the sound speed $c_s$ equals the speed of light, i.e.,
$c_s = 1$ \cite{Chen:2006nt}.


Now, to obtain the analytical relations for $n_{s}$ and $r$, we need to calculate the scalar and tensor power spectra. As previously stated, our model includes multiple stages of inflation driven by each scalar field. Therefore, using the $\delta N$ formalism \cite{Sasaki1995}, we can calculate the power spectrum of the curvature perturbation $\mathcal{R}$, which directly depends on $Q^{b}$, as follows.
\begin{eqnarray}\label{PR}
\mathcal{P}_{\mathcal{R}}&=&\Big(\frac{H}{2 \pi}\Big)^2 \frac{\delta^{ab}}{f} \frac{\partial N}{\partial \phi^{a}}\frac{\partial N}{\partial \phi^{b}}\\ \nonumber &=&\Big(\frac{H}{2 \pi}\Big)^2 \Big[\frac{V^2 f^3}{V^{,a}V_{,a}(2f-1)^2 }\Big] \simeq \frac{H^2}{8 \pi^2 \epsilon_{H}}.
\end{eqnarray}
We used Eqs. \eqref{dphidN} and \eqref{epsilonH} in the final equality. 
Subsequently, similar to single-field inflation, one can compute the spectral index $n_{s}$ as follows\footnote{This result is consistent with the spectral index of the curvature perturbations obtained by \cite{Sasaki1995}
\begin{equation}
n_{s}-1=-2 \epsilon_{H}+\frac{1}{6  \pi^2 \mathcal{P}_{\mathcal{R}}} (M^{2})^{ab}_{\rm SR} \frac{\partial N}{\partial \phi^{a}} \frac{\partial N}{\partial \phi^{b}}, 
\end{equation}
where in the slow-roll (SR) limit, the effective mass is
\begin{equation}\label{massSR}
\nonumber {(M^2)}^{ab}_{\rm SR} \approx \tilde{V}^{;ab}- {{\mathbb{R}^{a}}_{cd}}^{b} \dot{\phi}^{c}\dot{\phi}^{d}-3\dot{\phi}^{a} \dot{\phi}^{b}\Big(\frac{1}{2f-1}\Big),
\end{equation}
in which, in two-field scenario, the Riemann tensor can be expressed as
 \begin{equation}
 \mathbb{R}_{abcd}= \frac{f^2 \mathbb{R}}{2} \Big(\delta_{ac}\delta_{bd}-\delta_{ad}\delta_{bc}\Big),
 \end{equation}
 where $\mathbb{R}=\delta^{nm}(f_{,n}f_{,m}-ff_{,nm})/{f^3}$ is Ricci scalar.}:
 \begin{equation}\label{nsfunction}
 n_{s}-1 \equiv \frac{d \ln \mathcal{P}_{\mathcal{R}}}{d \ln k} \simeq -2 \epsilon_{H}- \eta_{H},
 \end{equation}
 where
 \begin{equation}\label{etafunction}
 \eta_{H}=\frac{\dot{\epsilon_{H}}}{H \epsilon_{H}}=\frac{d \phi^{a}}{dN}\frac{ {\epsilon_{H}}_{,a}}{\epsilon_{H}}.
 \end{equation}
Additionally, the tensor-to-scalar ratio $r$ is another key observable in inflationary models. The tensor spectrum follows the standard formula for the single-field inflation, which is given by
\begin{equation}\label{PT}
\mathcal{P}_{T}=\frac{2 H^2}{\pi^2} \simeq \frac{2}{3 \pi^2} f V.
\end{equation}
 \begin{figure}
\centering
	\includegraphics[width=0.56\textwidth]{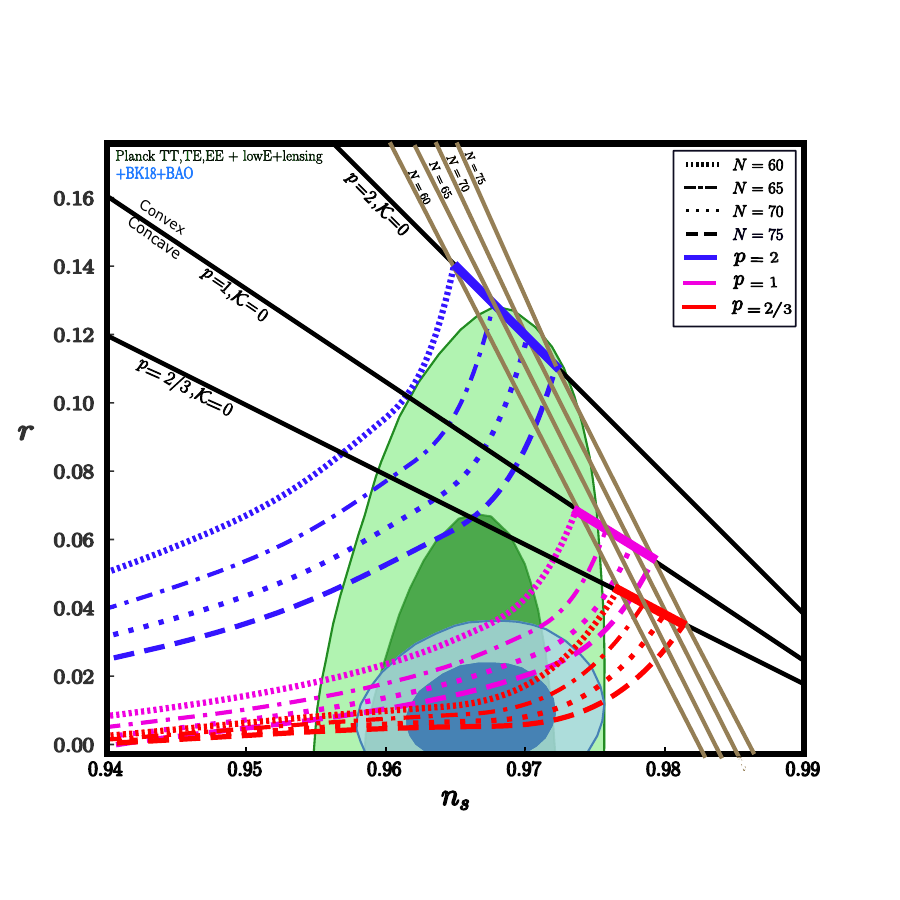}
	\caption{ The tensor-to-scalar ratio as a function of the spectral index for $2/3 \leq p \leq 2$. For different values of $N_{\rm end}$ (i.e., $N_{\rm end}=60$, $65$, $70$, and $75$), the number of e-folds for the first scalar fields in the first phase is $N_{1}=57$, $N_{1}=62$, $N_{1}=67$, and $N_{1}=72$, respectively. The number of e-folds for the second scalar fields in the second phase is $N_{2}=3$ for all cases.}
	\label{fig3}
\end{figure}
Therefore, by using Eqs. \eqref{PT} and \eqref{PR}, the tensor-to-scalar ratio is obtained as
\begin{equation}\label{rfunction}
r \equiv \frac{\mathcal{P}_{T}}{\mathcal{P}_{\mathcal{R}}}\simeq \frac{8 (2f-1)^2}{V^2 f^3} \delta^{ab} V_{,a}V_{,b} \simeq 16 \epsilon_{H}.
\end{equation}
Until now, our analysis has been general and applicable to any number of fields. We will now explicitly consider the case where only two scalar fields are present. Therefore, by combining Eqs. \eqref{epsilonH}, \eqref{etafunction}, \eqref{nsfunction}, and \eqref{rfunction} together with Eqs. \eqref{anaphi1} and \eqref{anaphi2}, the spectral index $n_{s}$ and the tensor-to-scalar ratio $r$ at the CMB scale at $N=0$, are expressed as
 \begin{eqnarray}\label{nsA}
\nonumber  &n_{s}&-1\simeq -\frac{1}{N_{1}}-\frac{3 p}{2 N_{1}}\Big[(2pN_{1})^{p} \beta^2 \frac{(6+p)}{12(1+p)}+(2 pN_{1})^{\frac{3p}{2}} \beta^{3}\\
\nonumber  &\times & \frac{3(2+p)}{4(2+3p)}+(2 p N_{1})^{2p} \beta^{4} \frac{(42+p(107+4p(22+7p)))}{12(1+p)^2(1+2p)}\\
 &+& \mathcal{O}\Big(\beta^{5}, \frac{\mu_{2}}{\mu_{1} }N_{2}^{\frac{p}{2}}\Big) 
 \Big] 
 \end{eqnarray}
 and 
 \begin{eqnarray}\label{rA}
\nonumber  r &\simeq & \frac{4p}{N_{1}} \Big[1-\frac{2p N_{1}^{\frac{p}{2}}}{2} \beta
 -(2pN_{1})^{p} \beta^{2} \frac{(1+2p)}{4(1+p)}\\
\nonumber  &-&2(2pN_{1})^{\frac{3p}{2}} \beta^{3} \frac{(4+p(16+9p))}{16(1+p)(2+3p)}-(2pN_{1})^{2p} \beta^{4}\\
\nonumber  &\times & \frac{(2-p(p(p(3+10p)-17)-13))}{16 (1+p)^2(2+p(7+6p))}\\
 &+& \mathcal{O}\Big(\beta^{5}, \frac{\mu_{2}}{\mu_{1} }N_{2}^{\frac{p}{2}}\Big) 
 \Big]. 
 \end{eqnarray}
 

It is evident that the $\beta$ corrections provide a clear improvement in the values of $r$ and $n_{s}$ for standard two-field inflation. Specifically, due to the positive nature of $\beta$, all corrections have the potential to decrease $n_{s}$ and $r$ compared to standard two-field inflation, bringing them closer to current observational constraints. For instance, to be compatible with large-scale CMB observations from the Planck
TT,TE,EE+lowE+lensing+BK15+BAO results at the pivot scale \cite{aghanim2020planck},
\begin{equation}
0.956<n_{s}<0.978, \hspace{0.3cm} r(k_{*})\leq 0.066 \hspace{0.3cm} \text{at} \hspace{0.3cm} 95 \% \text{C.L}
\end{equation}
the $\beta$ related to the potential model with $p=2$ must be in the range $[0.00174332, 0.00207482)$ for $N_{1}=72$ and $N_{2}=3$. It should be noted that this model was previously ruled out by the above bound in the standard two-field inflation by the $r = 4p/N_{\rm end}$ relation, even for $N_{\rm end}=75$ \cite{Easther:2005zr,Wenren:2014,starobinsky1985multicomponent,lyth1999particle}
(See the solid black straight lines in Fig. \ref{fig3}).

We have also provided the $\beta$ range for other models in accordance with Planck, along with joint constraints from BAO and BK18 \cite{ade2021improved}, as collected in Tab. \ref{tab1}. Here, the number of e-folds for the second scalar fields in the second phase is $N_{2}=3$ for all cases.

\begin{table}[h!]
\centering
\scriptsize
\begin{tabular}{c| c| c} 
 \hline
Model & $\#$ e-foldings & Range \\ [1ex] 
 \hline
        & 60 & $0.157878\leq\beta<0.201486$   \\
$p = 2/3$ & 65 & $0.164147<\beta< 0.203895$      \\
        & 70 & $0.173103\leq\beta< 0.209411$   \\       
        & 75 & $0.178119<\beta< 0.212407$     \\
  \hline
        & 60 & $0.0500187<\beta< 0.0673508$    \\
$p = 1$   & 65 & $0.0525710\leq\beta< 0.0704950$ \\
        & 70 & $0.0605319<\beta< 0.0783060$   \\
        & 75 & $0.0593465 <\beta< 0.0724056$   \\
  \hline
\end{tabular}
\caption{The range of $\beta$ for inflationary parameters in two-field model. }
\label{tab1}
\end{table}

To clearly illustrate the results presented in Tab. \ref{tab1}, we have depicted the tensor-to-scalar ratio as a function of the spectral index using Eqs. \eqref{nsA} and \eqref{rA}, along with observational constraints from the Planck 2018 data, BICEP/Keck (BK15 \cite{aghanim2020planck} and BK18 \cite{ade2021improved}) data, and BAO data in Fig. \ref{fig3}. Notice that the BK18 analysis improved the 95$\%$ confidence constraint from BK15, reducing it from $r_{0.05} < 0.066$ to $r_{0.05} < 0.036$. Additionally, the BK18 simulations produced a median 95 $\%$ upper limit of $r_{0.05} < 0.019$.

As can seen, the model $p=2$ with $N_{\rm end}=75$ is located in the region determined by the BK15, whereas both models $p=1$ and $p=2/3$ fall entirely within the region defined by the BK18 results. Specifically, by adjusting both $n_{s}$ and $r$ values downward, a model like $p=2$, previously excluded by BK15 in standard two-field inflation, now falls within the observational range defined by BK15. Additionally, the predicted values of ${n_{s}, r}$ in the $p=2/3$ model, with $\beta$ corrections, fall entirely within the region determined by the BK18 results \cite{ade2021improved}, unlike in the standard two-field model with $\mathcal{K}=0$, where they are ruled out. 

Before leaving this section, let us compare  the analytical and numerical results for $\{n_{s},r\}$ in the $p=2$ case\footnote{For another model like $p=2/3$, to manage the inflationary transition from the first rolling field toward its minimum to the second field, one needs to impose additional potential terms such as $\mu_{3}^{2} \phi_{1}^2$ (where $\mu_{3} \sim \mu_{2} \ll \mu_{1}$) to create an artificial minimum for the first field potential. However, finding the mass of this term requires fine-tuning. Therefore, we restrict ourselves to numerically evaluating the inflationary parameters for the model $p=2$, which has a definite minimum.}. As shown in Tab. \ref{tab2}, the analytical and numerical results exhibit good agreement for various values of the number of e-foldings.
\begin{table}[h!]
\centering
\scriptsize
\begin{tabular}{c|c|c|c|c|c} 
 \hline
$\beta$& $\#$ e-foldings &$r_{\text{Numeric}}$ & $r_{\text{Analytic}}$ & $n_{s_{\text{Numeric}}}$& $n_{s_{\text{Analytic}}}$ \\ [1ex] 
 \hline
0.0017 &  60    & 0.105  &  0.103  & 0.9620  & 0.9614 \\
 \hline
0.0023 &  65    & 0.074  &  0.072  & 0.9591  & 0.9580 \\ 
 \hline
0.0021 &  70    & 0.069  &  0.068  & 0.9624  & 0.9615 \\
 \hline 
0.002 &  75    & 0.062  &  0.0615  & 0.9645  & 0.9636 \\  [1ex] 
 \hline
\end{tabular}
\caption{ Comparing between the analytical and numerical results for inflationary parameters in the quadratic potential with $p=2$. The second inflationary phase takes place $N_{2}=3$ for all cases. Note that the analytical results were obtained by substituting the values of $N_{1}$, $N_{2}$, and $\beta$ into Eqs. \eqref{nsA} and \eqref{rA}.} 
\label{tab2}
\end{table}


\section{Conclusion}\label{Sec3}
 
In this study, we introduced a specific subset of generalized multi-field inflation that incorporates a coupling term between the potential function and the kinetic term, represented as $XV$. The presence of this coupling term significantly impacts the slow-roll parameters and consequently leads to notable changes in our predictions for the scalar spectral index and the tensor-to-scalar ratio at CMB scales. 

To be more precise, due to the hierarchy of masses, we have analytically demonstrated that $n_{s}$ and $r$ depend on the number of e-folds $N_{1}$ during the first inflationary phase and the coupling correction. In fact, these modifications cause both observational quantities, $n_{s}$ and $r$, to shift to smaller values compared to standard multi-field inflation. Additionally, we examined the consistency between these analytical relations and the numerical data for the quadratic potential model with $p=2$.

In summary, incorporating a non-separable coupling between the kinetic and potential terms in both single-field and multi-field inflation setups enables us to align the parameters $n_{s}$ and $r$ with the recent Planck constraints. 

In light of the current investigations, it would be interesting to examine the non-Gaussianity of our model and explore the possibility of addressing both the Hubble parameter $H_{0}$ and the growth of structure parameter $S_{8}$ tensions simultaneously, as observed in the single-field setup.

\section*{Acknowledgments}

We gratefully acknowledge Hassan Firouzjahi and Alireza Talebian for useful discussions and correspondence. We also thank Shahram Khosravi and Shaghyegh Aalaei for their preliminary discussions on this topic.  
FF and PC are supported by the NSRF via the Program Management Unit for Human Resources $\&$ Institutional Development, Research and Innovation [grant number B13F670063].



\bibliography{multi-field-inflation}

\end{document}